\documentclass[letterpaper, 10 pt, conference]{ieeeconf}  % Comment this line out if you need a4paper

\IEEEoverridecommandlockouts                              % This command is only needed if
\usepackage{cite}
\usepackage{amsmath,amssymb,amsfonts, lipsum, mathtools, cuted}
\usepackage{graphicx}
\usepackage{changes}
\definechangesauthor[name={Farhad Ghanipoor}, color=blue]{FG}

\usepackage{ntheorem}
\usepackage{thmtools}
\newtheorem{theorem}{Theorem}

\newtheorem{lemma}{Lemma}
\newtheorem{definition}{Definition}
\newtheorem{propos}{Proposition}
\newtheorem{assumption}{Assumption}
\newtheorem{remark}{Remark}
\newtheorem{problem}{Problem}

\allowdisplaybreaks

\title{\LARGE \bf
Ultra Local Nonlinear Unknown Input Observers\\ for Robust Fault Reconstruction
}

\author{Farhad Ghanipoor$^{1}$, Carlos Murguia$^{1}$, Peyman Mohajerin Esfahani$^{2}$, and Nathan van de Wouw$^{1}$
	\thanks{$^{1}$ Farhad Ghanipoor, Carlos Murguia,  and Nathan van de Wouw are with the Mechanical Engineering Department, Eindhoven University of Technology, Eindhoven, The Netherlands.  {Emails: \tt\small f.ghanipoor@tue.nl}, {\tt\small C.G.Murguia@tue.nl}, and {\tt\small N.v.d.Wouw@tue.nl}.}%
	\thanks{$^{2}$ Peyman Mohajerin Esfahani is with the Delft Center for Systems and Control, Delft University of Technology, Delft, The Netherlands. {Email: \tt\small P.MohajerinEsfahani@tudelft.nl}.}%
}

\begin{document}

	\maketitle
	\thispagestyle{empty}
	\pagestyle{empty}

	%%%%%%%%%%%%%%%%%%%%%%%%%%%%%%%%%%%%%%%%%%%%%%%%%%%%%%%%%%%%%%%%%%%%%%%%%%%%%%%%
\begin{abstract}
In this paper, we present a methodology for actuator and sensor fault estimation in nonlinear systems. The method consists in augmenting the system dynamics with an approximated ultra-local model (a finite chain of integrators) for the fault vector and constructing a Nonlinear Unknown Input Observer (NUIO) for the augmented dynamics. Then, fault reconstruction is reformulated as a robust state estimation problem in the augmented state (true state plus fault-related state). We provide sufficient conditions that guarantee the existence of the observer and stability of the estimation error dynamics (asymptotic stability of the origin in the absence of faults and ISS guarantees in the faulty case). Then, we cast the synthesis of observer gains as a semidefinite program where we minimize the $\mathcal{L}_2$-gain from the model mismatch induced by the approximated fault model to the fault estimation error. Finally, simulations are given to illustrate the performance of the proposed methodology.
\end{abstract}

	%%%%%%%%%%%%%%%%%%%%%%%%%%%%%%%%%%%%%%%%%%%%%%%%%%%%%%%%%%%%%%%%%%%%%%%%%%%%%%%%
	\section{INTRODUCTION} \label{sec: introduction}
	
The increasing expectation of product quality and cost-efficient operation of engineered systems has led to an increasing demand on process reliability. This can be achieved through predictive maintenance technology, for which methods and techniques for fault detection (is a fault occurring?), isolation (what is the fault source?), and estimation (how large the fault is?) are fundamental ingredients. Fault Detection and Isolation (FDI) has been an active field of research for many years (see, for instance, \cite{ding2008model}\nocite{zhang2002robust}\nocite{hwang2009survey}\nocite{Moradmand2020}\nocite{chen2012robust}\nocite{narasimhan2008new}\nocite{esfahani2015tractable}-\cite{de2001geometric}, and references therein). However, most solutions have been developed for engineered systems with linear dynamics, when in practice many of them are highly nonlinear in nature (e.g., robotics, power, transportation, water, and manufacturing). The standard path to tackle nonlinear behaviour is to obtain approximated linear models of the system and then apply linear FDI techniques. Even though this might provide insight about the presence and source of the fault, using linear models on nonlinear systems often leads to high false negatives/positives since linear FDI schemes struggle to distinguish between true faults and model mismatches. The latter has motivated the development of nonlinear FDI schemes built around different ideas \cite{chen2012robust}\nocite{narasimhan2008new}\nocite{esfahani2015tractable}-\cite{de2001geometric}, e.g., observer-based \cite{chen2012robust},\cite{narasimhan2008new}, optimization-based \cite{esfahani2015tractable}, and differential-geometric methods \cite{de2001geometric}. Most of these results provide elegant FDI solutions; however, they impose strong hard to verify conditions on the system dynamics -- mainly fault-output decoupling by filtering or observability with respect to different subsets of outputs -- which significantly limits their applicability. Moreover, despite the great value of existing FDI schemes, what is mainly required for predictive maintenance is fault estimation. That is, we not only need to know the presence and source of the fault, but also its severity. If the fault is small and/or slowly growing in magnitude (slow compared to the system time-scale), predictions of the fault severity can be performed and predictive maintenance can be scheduled accordingly. The latter is only possible if we can reconstruct fault signals (at least their magnitude) with the available information (inputs, outputs, and system models). Note that by estimating the fault vector, we automatically have detection and isolation by looking at the support (the nonzero entries) of the estimated vector.

As it is the case for FDI, fault estimation results are fairly mature for linear systems (see, e.g., \cite{liu2013sensor,liu2012fuzzy} for results on linear stochastic and switchings systems) and still developing for nonlinear dynamics. In \cite{Adaptive}, the authors address the problem for nonlinear systems with uniformly Lipschitz nonlinearities, process faults only (i.e., no sensor faults), and assume the so-called matching condition (the rank of the fault distribution matrix is invariant under left multiplication by the output matrix). For this configuration, they provide an adaptive filter capable of approximately reconstructing the actuator fault vector. The matching condition, however, is a strong assumption, it makes the problem tractable but significantly reduces the class of system for which results are applicable. In \cite{Phan2021}, a fault estimation scheme is introduced for both sensor and process faults using Nonlinear Unknown Input Observers (NUIO), adaptive Radial Basis Function Neural Networks (RBFNN), and assuming the matching condition. They prove their scheme provides boundedness of fault estimation errors.

The authors in \cite{zhu2015fault} do not assume the matching condition. They consider Lipschitz nonlinearities, simultaneous sensor and process faults, and assume a standard fault observability condition \cite{Patton_Input_Observability} on the linear part of the dynamics. They tackle the problem using the notion of intermediate observers, which consists on having two dedicated observers, one that estimates the fault and the other the state. Their scheme guarantees bounded fault estimation errors. In \cite{van2020multiple}, simultaneous additive and multiplicative process faults are considered. They address the fault estimation problem by decoupling process nonlinearities and perturbations from the estimation filter dynamics, and using regression techniques to approximately reconstruct fault signals. Decoupling nonlinearities leads to linear filters for which linear techniques can be used to reconstruct fault signals. However, decoupling conditions impose strong assumptions on the system dynamics, which significantly limits the applicability of these results.

We remark that the above mentioned results for nonlinear systems guarantee approximate reconstruction of fault vectors only, i.e., they ensure bounded estimation errors which, if small enough, still lead to a fairly good estimate of the true fault. Not having internal models of fault signals makes it challenging to enforce zero error fault estimation. To this end, we propose a fault estimation scheme, for process and sensor faults, uniformly Lipschitz nonlinearities, an without assuming the matching condition, that incorporates an approximated internal model of the fault vector. We use the notion of \emph{ultra-local models} \cite{Flies_Ultra_Local}-\cite{sira2018active} -- which refers to a class of phenomenological models that are only valid for very short time intervals -- to characterize internal fault models. We then extend the system dynamics with an approximated internal fault model and construct a NUIO to jointly reconstruct faults and states. The fault estimation problem is re-formulated as a robust state estimation problem in the augmented state (true state plus fault-related state). We provide sufficient conditions that guarantee the existence of the observer and stability of the estimation error dynamics (asymptotic stability of the origin in the absence of faults and ISS guarantees in the faulty case). Then, we cast the synthesis of observer gains as a semidefinite program where we minimize the $\mathcal{L}_2$-gain from the model mismatch induced by the approximated fault model to the fault estimation error.

	\textbf{Notation:}
	The symbol $\mathbb{R}^{+}$ denotes the set of nonnegative real numbers. The $n \times n$ identity matrix is denoted by $I_n$	or simply $I$ if $n$ is clear
	from the context. Similarly, $n \times m$ matrices composed of
	only zeros are denoted by $\boldsymbol{0}_{n \times m}$ or simply $\boldsymbol{0}$ when their dimensions are clear. For positive definite (semi-definite) matrices, we use the notation $P \succ 0$ $(P \succeq 0)$. For negative definite (semi-definite) matrices, we use the notation $P \prec 0$ $(P \preceq 0)$. The notation $\text{col}[x_1, \ldots , x_n]$ stands for the column vector composed of the elements $x_1,\ldots,x_n$. This notation is also used when the components $x_i$ are vectors. The $\ell_2$ vector norm (Euclidean norm) and the matrix norm induced by the $\ell_2$ vector norm are both denoted as $||\cdot||$. We use $\mathcal{L}_2(0,T)$ (or simply $\mathcal{L}_2$) to denote vector-valued functions $z:[0,T] \to \mathbb{R}^{k}$ satisfying $\int_{0}^{T}\|z(t)\|^{2} d t<\infty$. For a vector-valued signal $f(t)$ defined for all $t \geq 0$, $||f(t)||_\infty := \sup_{t \geq 0} ||f(t)||$. For a differentiable function $W: \mathbb{R}^{n}  \to \mathbb{R}$ we denote by $\frac{\partial W}{\partial e}$ the row-vector of partial derivatives and by $\dot{W}(e)$ the total derivative of $W(e)$ with respect to time.

%	norm $\|A\| = \text{sup}_{\|x\| \leq 1} \|A x\| $.
	
\section{Problem Formulation} \label{sec: problem formulation}
Consider the nonlinear system
	\begin{equation} \label{eq: sys_no_last_sensor}
		 \left\{\begin{aligned}
			\dot{x} (t) =& A x(t)+ B u(t)+ S g(Vx(t), u(t)) +F_x f(t),\\
			y(t)=& C x(t) + F_y f(t),
		\end{aligned}\right.
	\end{equation}
where $t \in \mathbb{R}^{+}$, $x \in {\mathbb{R}^{n}}$, $y \in {\mathbb{R}^{{m}}}$, and $u \in {\mathbb{R}^{{l}}}$ are time, state, measured output and known input vectors, respectively, ${n},{m}, l \in \mathbb{N}$, matrices $(A, B, S, V, C, F_x, F_y)$ of appropriate dimensions, and $g: \mathbb{R}^{n_v} \times \mathbb{R}^{l} \to \mathbb{R}^{n_g}$ is a nonlinear vector field. Function $f: \mathbb{R}^{+}  \to \mathbb{R}^{n_f}$ denotes the unknown fault vector, which contains both process and sensor faults, $f_{x}(t)$ and $f_{y}(t)$, respectively, $f(t):=\left[\begin{array}{ll}f_{x}^{T}(t) & f_{y}^{T}(t)\end{array}\right]^{T}$. Matrix $F_x$ denotes the process fault distribution matrix while matrix $F_y$ represents the contribution of the fault signal to sensor measurements. Matrix $S$ indicates in which equation(s) the nonlinearity appears explicitly, and $V$ is used to indicate which states play a role in the nonlinearity. We often omit implicit time dependencies for notation simplicity. 	

\begin{assumption}\emph{\textbf{(Globally Lipschitz Nonlinearity)}} Vector field $g(Vx,u)$ in \eqref{eq: sys_no_last_sensor} is globally Lipschitz uniformly in $u(t)$ and $t$, i.e., there exists a known positive constant $\alpha$ satisfying
		\begin{equation}
			\|g(V\hat{x},u,t)-g(V{x},u,t)\| \leq \alpha \|V (\hat{x}-x)\|,
			\label{eq: lipschitz}
		\end{equation}
for all $x,\hat{x} \in {\mathbb{R}^{n}}$, $u \in {\mathbb{R}^{l}}$, and $t \in {\mathbb{R}^{+}}$.
	\end{assumption}

\begin{assumption}\emph{\textbf{($\mathcal{C}^r$ Fault Vector)}} The fault vector $f(t)$ in \eqref{eq: sys_no_last_sensor} is $r$ times differentiable, i.e., the time derivatives ${f}^{(1)}(t)$, ${f}^{(2)}(t)$, ... ,$f^{(r)}(t)$ exist and are continuous, and $f^{(r)}(t)$ is uniformly bounded.
	\end{assumption}

\subsection{Ultra Local Fault Model}

Under Assumption 2, we can write the linear approximated model $\zeta_i(t) = f^{(i-1)}(t)$, $i=1,2,...,r-1$, $\zeta_1^{(r)}(t) = \mathbf{0}$, for the fault vector as
\begin{equation}\label{fault_model}
\left\{\begin{aligned}
  \dot{\zeta}_1 &= \zeta_2,\\
  \dot{\zeta}_2 &= \zeta_3,\\
  \vdots \\
  \dot{\zeta}_{r-1} &= \zeta_r,\\
  \dot{\zeta}_{r} &= \mathbf{0},\\
  f &= \zeta_1.
\end{aligned}\right.
\end{equation}
Note that this model corresponds to an entry-wise $r$-th order Taylor time-polynomial approximation at time $t$ of $f(t)$. The accuracy of the approximated model increases as $f^{(r)}(t)$ goes to zero (entry-wise) and it is exact for $f^{(r)}(t) = \mathbf{0}$. Model \eqref{fault_model} is used to construct an observer that ultra-locally \cite{Flies_Ultra_Local}-\cite{sira2018active} acts as a self-updating polynomial spline approximating the actual value of the fault. To design such an observer, we extend the system state, $x(t)$, with the states of the fault model, $\zeta_i(t)$, $i=1,2,...,r$, and augment the system dynamics in \eqref{eq: sys_no_last_sensor} with \eqref{fault_model}. We then design a Nonlinear Unknown Input Observer (NUIO) for the augmented system to simultaneously estimate $x(t)$ and $f(t)$. We remark that the number of the faults derivatives, $r$, added to the approximated model \eqref{fault_model} is problem-dependent and an optimal selection would depend on the frequency characteristics of the fault. Increasing $r$ results in higher dimensional augmented dynamics, and thus high dimensional observers as well. However, having larger observers also provides more degrees of freedom for optimal synthesis.

\subsection{Augmented Dynamics}
Define the augmented state $x_a:=\text{col}[x,f,\dot{f},\ldots,f^{(r-1)}]$ and write the augmented dynamics using \eqref{eq: sys_no_last_sensor} and \eqref{fault_model} as
		\begin{subequations} 		\label{eq: augmented}
	\begin{equation}
		\begin{aligned}
			\left\{\begin{aligned}
				\dot{x}_{a} &=A_{a} x_{a}+B_a u+ S_a g(V_a x_a, u) +D_{a} f^{(r)},\\
				y &= C_{a} x_{a},
			\end{aligned}\right.
		\end{aligned}
		\label{eq: augmented_system}
	\end{equation}
	where
	\begin{equation}
		\begin{aligned}
			A_{a} &:=\left[\begin{array}{ccccc}
				A & F_x & \boldsymbol{0} & \ldots & \boldsymbol{0} \\
				\boldsymbol{0} & \boldsymbol{0} & I_{n_f} & \ldots & \boldsymbol{0} \\
				\vdots & \vdots & \vdots & \ddots & \vdots \\
				\boldsymbol{0} & \boldsymbol{0} & \boldsymbol{0} & \ldots & I_{n_f} \\
				\boldsymbol{0} & \boldsymbol{0} & \boldsymbol{0} & \ldots & \boldsymbol{0}
			\end{array}\right], B_a :=	\left[\begin{array}{cc}
				B \\
				\boldsymbol{0}
			\end{array}\right],\\
			S_a &:= \left[\begin{array}{cc}
				S^T &
				\boldsymbol{0}
			\end{array}\right]^T,
			V_a := [\begin{array}[]{ll}V & \boldsymbol{0} \end{array}],\\
			D_a &:= \left[\begin{array}{cc}
				\boldsymbol{0} &
				I_{n_f}
			\end{array}\right]^T, C_a := \left[\begin{array}{lll}	C & F_y	& \boldsymbol{0} \end{array}\right].\\
		\end{aligned}
		\label{eq: augmented_matrices}
	\end{equation}
			\end{subequations}

\subsection{Joint State-Fault Nonlinear Observer}
	We propose the following nonlinear unknown input observer to estimate the augmented state ${x}_a$
	\begin{subequations} \label{eq: nuio}
	\begin{equation}
		\begin{aligned}
			\dot{z} &=N z+G u+L y+M S_a g(V_a\hat{x}_a+ J(y - C_a \hat{x}_a),u), \\
			\hat{x}_a &=z-E y,
		\end{aligned}
		\label{eq: nuio_sys}
	\end{equation}
with observer state $z \in {\mathbb{R}^{n_z}}$ ($n_z = n+rn_f$), estimate of the extended estate $\hat{x}_a$, and matrices $(N,G,L,M)$ defined as
	\begin{equation}
		\begin{aligned}
			N &:=M A_a-K C_a,
			&G&:=M B_a, \\
			L &:=K(I+C_a E)-M A_a E,
			&M &:=I+E C_a.
		\end{aligned}
		\label{eq: nuio_matrices}
	\end{equation}
	\end{subequations}
Matrices $E$, $K$, and $J$ are observer gains to be designed. Note that the fault estimate $\hat{f}(t)$ is given by $\hat{f} = \bar{C}\hat{x}_a$ with
		\begin{equation}
\bar{C} := {\left[\begin{array}{lll}
		\boldsymbol{0}_{n_f \times n} & I_{n_f} & \boldsymbol{0}_{n_f \times n_f(r-1)}
	\end{array}\right]}.
\label{eq: c_bar}
		\end{equation}
Define the estimation error as
	\begin{equation*}
		e:=\hat{x}_a-x_a=z-x_a-E y=z-M x_a.
	\end{equation*}
	Then, the estimation error dynamics is given by
	\begin{equation}
		\begin{aligned}
			\dot{e}&= N e+(N M+L C_a-M A_a) x_a+(G-M B_a) u \\
			&+M S_a\big(g(V_a\hat{x}_a + J (y - C_a \hat{x}_a), u)-g(V_a {x}_a, u)\big) \\
			&-M D_a f^{(r)}.
		\end{aligned}
		\label{eq: observer_dynamic}
	\end{equation}
Given the algebraic relations in \eqref{eq: nuio_matrices}, it can be verify that $G-MB_a = \mathbf{0}$ and $N M+L C_a-M A_a = \mathbf{0}$. Therefore, \eqref{eq: observer_dynamic} can be written as
	\begin{equation}
		\begin{aligned}
			\dot{e} &= Ne + M S_a \delta g- M D_a f^{(r)},
		\end{aligned}
		\label{eq: observer_dynamics_1}
	\end{equation}
where $\delta g := g(V_a\hat{x}_a + J (y - C_a \hat{x}_a), u)-g(V_a {x}_a, u)$. We have now all the machinery required to state the problem we seek to solve.

\begin{problem}\emph{\textbf{(Fault Reconstruction)}} Consider the nonlinear system \eqref{eq: sys_no_last_sensor} with known input and output signals, $u(t)$ and $y(t)$, and let Assumption 1 and Assumption 2 be satisfied. Further consider the approximated internal fault model \eqref{fault_model}, the augmented dynamics \eqref{eq: augmented_system}, and the observer \eqref{eq: nuio}. Design the observer gain matrices $(E,K,J)$ so that: \textbf{\emph{1)}} all trajectories of the estimation error dynamics \eqref{eq: observer_dynamics_1} exists and are globally ultimately bounded uniformly in $t \geq 0$\emph{;} \emph{\textbf{2)}} $\int_{0}^{T}||\bar{C}e(t)||^{2} d t \leq c \int_{0}^{T}\|f^{(r)}(t) \|^{2} dt$, for some known $c > 0$, all $T \in {\mathbb{R}^{+}}$, and $\bar{C}$ in \eqref{eq: c_bar}; and \emph{\textbf{3)}} for $f^{(r)}(t) = \mathbf{0}$, $t \geq 0$, $\lim_{t \rightarrow \infty} ||e(t)|| = 0$.
\end{problem}

Under Assumption 1 and Assumption 2, Problem 1 amounts to finding a fault estimator that guarantees a bounded estimation error, $e(t)$, that $e(t)$ goes to zero asymptotically if $f^{(r)}(t) = \mathbf{0}$, and that the $\mathcal{L}_2(0,T)$ norm of $\bar{C}e(t)$ (the fault estimation error) is upper bounded by that of $cf^{(r)}(t)$, for some $c>0$.

\section{Solution to Problem 1} \label{sec: lmi-based-design}

\subsection{ISS Estimation Error Dynamics}

In this section, we derive LMI conditions for designing the matrices $E, K$ and $J$ of the augmented system observer in \eqref{eq: nuio}. As a stepping stone, we present a sufficient condition for asymptotic stability of the origin of the estimation error dynamics when there is no fault, then we prove boundedness of the estimation error in the presence of faults using input-to-state stability ideas \cite{khalil2002nonlinear}.

	\begin{definition} \emph{\textbf{(Input-to-State Stability~{\cite[Def. 4.7]{khalil2002nonlinear}})}} The error dynamics \eqref{eq: observer_dynamics_1} is said to be Input-to-State Stable (ISS) if there exist a class $\mathcal{K} \mathcal{L}$ function $\beta(\cdot)$ and a class $\mathcal{K}$ function $\gamma(\cdot)$ such that for any initial estimation error $e(t_0)$ and any bounded $f^{(r)}$, the solution $e(t)$ of \eqref{eq: observer_dynamics_1} exists for all $t \geq t_{0}$ and satisfies
		\begin{equation}\label{ISS_def}
			\|e(t)\| \leq \beta\left(\left\|e\left(t_{0}\right)\right\|, t-t_{0}\right) + \gamma\big(\| f^{(r)}(t) \|_{\infty}\big).
		\end{equation}
	\end{definition}

Note that ISS of the error dynamics \eqref{eq: observer_dynamics_1} implies boundedness of the estimation error for bounded $f^{(r)}(t)$. This follow directly from $\eqref{ISS_def}$.

	\begin{lemma}\emph{\textbf{(ISS Lyapunov Function~{\cite[Thm. 4.19]{khalil2002nonlinear}})}}
		Consider the error dynamics \eqref{eq: observer_dynamics_1} and let $W(e)$ be a continuously differentiable function such that
		\begin{equation*}
			\alpha_{1}(\|e\|) \leq W(e) \leq \alpha_{2}(\|e\|),
		\end{equation*}
		\begin{equation*}
			\dot{W}(e) \leq-W_{3}(e), \quad \hspace{-1mm} \forall \hspace{1mm}\|e\| \geq \xi(\|f^{(r)}\|),
		\end{equation*}
		where $\alpha_{1}(\cdot)$ and $\alpha_{2}(\cdot)$ are class $\mathcal{K}_{\infty}$ functions, $\xi(\cdot)$ is a class $\mathcal{K}$ function, and $W_{3}$ is a continuous positive definite function. Then, the estimation error dynamics \eqref{eq: observer_dynamics_1} is ISS with gain $\gamma = \alpha_{1}^{-1}(\alpha_{2}(\xi))$.
		\label{lem: iss}
	\end{lemma}

Let $W(e):={e}^{T} P {e}$ be an ISS Lyapunov function candidate. Then, it follows from \eqref{eq: observer_dynamics_1} and the Lipschitz condition in \eqref{eq: lipschitz} (i.e., $\|\delta g\| \leq \alpha \|(V_a  - J C_a) {e}\|$) that
	\begin{equation}
		\begin{aligned} \dot{W}(e) &\leq e^{T} \Delta e  - 2 e^{T} P M D_a f^{(r)},
		\end{aligned}
		\label{eq: lyapanov}
	\end{equation}
	where $\Delta := N^{T} P+P N+\alpha P M S_a S_a^{T} M^{T} P+\alpha (V_a - J C_a)^{T} (V_a - J C_a)$. A complete derivation of inequality \eqref{eq: lyapanov} is given in Appendix \ref{ap: lyapanov}. Inequality \eqref{eq: lyapanov} implies the following
	\begin{equation}
		\begin{aligned} \dot{W}(e) \leq& -\lambda_{\min }(-\Delta) \|e\|^{2} + 2 \|e\| \|P M D_a\| \|f^{(r)}\| \\
			=& - (1-\theta) \lambda_{\min }(-\Delta) \|e\|^{2} - \theta \lambda_{\min }(-\Delta) \|e\|^{2} \\
			&+ 2 \|e\| \|P M D_a\| \|f^{(r)}\|,
		\end{aligned}
		\label{eq: iss_analysis}
	\end{equation}
	for any $\theta \in (0,1)$. Therefore, by \eqref{eq: iss_analysis} and Lemma \ref{lem: iss}, if $\Delta$ is negative definite, system \eqref{eq: observer_dynamics_1} is ISS with input $f^{(r)}$ and linear ISS-gain
	\begin{equation}
	\gamma(\|f^{(r)}\|)  = \frac{2 \|P M D_a\|}{\theta \lambda_{\min }(-\Delta)}\|f^{(r)}\|.
	\label{eq: gamma}
	\end{equation}
Based on the above discussion, the next proposition formalizes a LMI condition ($\Delta \prec 0$) that guarantees an ISS estimation error dynamics \eqref{eq: observer_dynamics_1}. Without loss of generality, for numerical tractability, we enforce $\Delta + \epsilon I \preceq 0$ for some small given $\epsilon >0$ instead of $\Delta \prec0$.
	
	\begin{propos}\emph{\textbf{(ISS Estimation Error Dynamics)}} Consider the error dynamics \eqref{eq: observer_dynamics_1} and suppose there exist matrices ${\mathbb{R}^{n_z \times n_z}} \ni P \succ0, R \in {\mathbb{R}^{n_z \times m}}, Q \in {\mathbb{R}^{n_z \times m}}$, and $J \in {\mathbb{R}^{n_v \times m}}$ satisfying the inequality
		\begin{equation}
			\left[\begin{array}{cc}
				X 	 +  \epsilon I & X_{12} \\
				X_{12}^{T} & -I
			\end{array}\right] \preceq 0,
			\label{eq: stability_lmi}
		\end{equation}
		for some given $\epsilon>0$ and matrices $X$ and $X_{12}$ defined as
		\color{black}
		\begin{equation}
			\begin{aligned}
				X :=& S_{11}+\alpha (V_a^{T} V_a -V_a^{T}  J  C_a - C_a^{T}  J^{T}  V_a), \\
				S_{11} := &A_{a}^{\top}  P +A_{a}^{\top} C_{a}^{\top} R^{\top}-C_{a}^{T}  Q^{T} + P A_{a} \\
				&+R C_{a} A_{a} -  Q C_{a}, \\
				X_{12} :=&\sqrt{\alpha}  {\left[\begin{array}{ll}
						(P+R C_a)S_a &  C_a^{T}  J^{T}
					\end{array}\right],}
			\end{aligned}
			\label{eq: X_X12}
		\end{equation}
%		Then, the estimation error dynamics \eqref{eq: observer_dynamics_1} is input-to-state stable with respect to the input $f^{(r)}$.
	with $\alpha$ from \eqref{eq: lipschitz}  and the remaining matrices in \eqref{eq: augmented_matrices}; then, the ISS-gain from input $f^{(r)}$ to the estimation error $e$ in \eqref{eq: observer_dynamics_1} is upper bounded by $ {2 \|P (I+E C_a) D_a\|}\epsilon^{-1}$ with $E = P^{^{-1}} R $.
		%			Therefore, regarding the stability of error dynamics \eqref{eq: observer_dynamics_1}, in the absence of fault, $e$ tends to zero asymptotically for any initial value of $e$. Moreover, $e$ is locally uniformly ultimately bounded in the fault presence when the derivative of the fault is bounded.
		
		\label{propos: stability}
	\end{propos}
	\emph{\textbf{Proof}}: We want to prove that $\Delta + \epsilon I \preceq 0$ is equivalent to \eqref{eq: stability_lmi}-\eqref{eq: X_X12}. Consider the expression for $\Delta$ in the text below \eqref{eq: lyapanov}. This expression is written in terms of the original observer gains $(E,K,J)$. Using this $\Delta$, \eqref{eq: nuio_matrices}, and Schur complements on $\Delta + \epsilon I \preceq 0$, we can write \eqref{eq: stability_lmi} with $X$ and $X_{12}$ in terms of the original observer gains $(E,K,J)$ as
	\begin{equation*}
		\begin{aligned}
			X &:= N^{T} P+P N +\alpha (V_a^{T} V_a -V_a^{T}  J  C_a - C_a^{T}  J^{T}  V_a), \\
			X_{12} &:= \sqrt{\alpha} {\left[\begin{array}{ll}
					P M S_a &  C_a^{T}  J^{T}
				\end{array}\right]},
		\end{aligned}
	\end{equation*}
where $N^{T} P+P N$ expands as
	\begin{equation*}
		A_{a}^{T}\left(I+E C_{a}\right)^{T} P-C_{a}^{T}  K P+P\left(I+E C_{a}\right) A_{a}-P K C_{a},
	\end{equation*}
	and $P M S_a$ becomes
	\begin{equation*}
		P S_{a}+P E C_{a} S_{a}.
	\end{equation*}
	Consider the following change of variables
	\begin{equation}
			R :=P E,  \qquad	Q :=P K.
			\label{eq: variable_change}
	\end{equation}
Applying \eqref{eq: variable_change} on the above expanded $X$ and $X_{12}$, the linear inequality \eqref{eq: stability_lmi}-\eqref{eq: X_X12} can be concluded. Clearly, $\Delta + \epsilon I \preceq 0$ implies $\lambda_{\min }(-\Delta) \geq \epsilon$. Then, using \eqref{eq: gamma}, we can conclude the bound on the ISS-gain.
	\hfill $\blacksquare$\\
	
\begin{remark}\emph{\textbf{(LMI Feasibility)}}
If $F_y$ is full row rank (i.e., there as many sensor faults, $f_y$, as sensors), the LMI in \eqref{eq: stability_lmi} is always infeasible (observability is lost) \cite{edwards2006comparison, edwards2000sliding, frank1989robust}. The standard practice to circumvent this issue is to assume $\text{rank}[F_y] < m$ \cite{edwards2006comparison, edwards2000sliding, frank1989robust}. Note that this is only a necessary condition for the estimator to exist, but it does not guarantee the LMI in \eqref{eq: stability_lmi} to be feasible (this has to be checked on a case by case basis).
 \label{rem: feasibility}
	\end{remark}
	
\begin{remark} \emph{\textbf{(Decoupling of Nonlinearities)}}
Note that it might be possible to cancel the effect of the nonlinearity $\delta g$ in the estimation error dynamics \eqref{eq: observer_dynamics_1} provided that we can select $M$ to satisfy $M S_a = \mathbf{0}$. This algebraic condition can be written in terms of the observer gain $E$ using the definition of $M$ in \eqref{eq: nuio_matrices} as follows
		\begin{equation}
			E C_a S_a = -S_a.
			\label{eq: no_nonlinear}
		\end{equation}
Equation \eqref{eq: no_nonlinear} has a solution $E$, if and only if matrix $C_a S_a$ has full column rank (details about this can be found in \cite{chen2006unknown}). Physically, the latter means there must be measurements from which the nonlinearity appears in the output dynamics. For the sake of generality, we do not assume this thought. We tackle the problem without imposing $M S_a = \mathbf{0}$ (so considering the nonlinear terms in the estimation error dynamics).
		\label{rem: no_nonlinear}
\end{remark}

Proposition 1 is used to enforce that all trajectories of the estimation error dynamics \eqref{eq: observer_dynamics_1} exist and are bounded for all $t \geq 0$. If the observer gains $(E,K,J)$ satisfy the ISS LMI in \eqref{eq: stability_lmi}, boundedness in guaranteed. This follows directly from Assumption 2 and Definition 1.

\subsection{$\mathcal{L}_2$ Performance Criteria}

To maximize the performance of the reconstruction scheme, we seek to minimize the effect of $f^{(r)}$ (treated as an external disturbance) on the estimation error dynamics \eqref{eq: observer_dynamics_1}. We could use the ISS formulation in Proposition 1 to cast an optimization problem where we minimize the ISS gain and treat the LMI in \eqref{eq: stability_lmi} as an optimization constraint. By doing so, we would be reducing the effect of $f^{(r)}$ on the complete vector of estimation errors $e$ (state, fault, and fault derivatives estimation errors). Note, however, that the purpose of the filter is to reconstruct faults only, so the performance in state estimation and in the error of higher order fault derivatives is not relevant. To this end, we seek to minimize the $\mathcal{L}_2$ gain from $f^{(r)}$ to the fault estimation error $\bar{C}e(t)$, with $\bar{C}$ in \eqref{eq: c_bar}. The $\mathcal{L}_2$ gain allows for an input-output ($f^{(r)} \rightarrow \bar{C}e(t)$) characterization of performance.

\begin{definition} \textbf{\emph{($\mathcal{L}_2$-gain~\cite{van19922})}}
We say that the estimation error dynamics \eqref{eq: observer_dynamics_1} with input $f^{(r)}$ and output $\bar{C}e(t)$ has a $\mathcal{L}_2$ gain less than or equal to $\lambda$ if the following is satisfied
		\begin{equation*}
			\int_{0}^{T}\|\bar{C}e(t)\|^{2} d t \leq \lambda^{2} \int_{0}^{T}\|f^{(r)}(t) \|^{2} d t,
		\end{equation*}
		for all $T \geq 0$ and $f^{(r)}(t) \in L_{2}(0,T)$.
		\label{def: l2}
	\end{definition}\

In the following lemma, we state a Lyapunov-based sufficient condition (the Hamilton-Jacobi inequality) for having a bounded $\mathcal{L}_2$ gain (see \cite[Thm. 5.5]{khalil2002nonlinear} and \cite[Thm. 2]{van19922} for further details).

	\begin{lemma} \emph{\textbf{($\mathcal{L}_2$-gain Inequality~{\cite[Thm. 5.5]{khalil2002nonlinear}})}}
		Consider \eqref{eq: observer_dynamics_1} and suppose there exists a continuously differentiable positive semi-definite function $W(e)$ satisfying
		\begin{equation}
			\begin{aligned}
				&\frac{\partial W}{\partial e} (Ne  +M S_a \delta g)+\frac{1}{2 \lambda^{2}} \frac{\partial W}{\partial e} M D_a D_a^{T} M^{T}\left(\frac{\partial W}{\partial e}\right)^{T}\\
				&\hspace{5mm}+\frac{1}{2} p^{T} p \leq 0, \\
			\end{aligned}
			\label{eq: hamilton-jacobi}
		\end{equation}
		with $p = \bar{C} e$ and $\bar{C}$ as in \eqref{eq: c_bar}. Then, the $\mathcal{L}_2$-gain from $f^{(r)}$ to the fault estimation error $p = \bar{C}e $ in \eqref{eq: observer_dynamics_1} is less than or equal to $\lambda$.
		\label{lem: l2_gain}
	\end{lemma}

	Based on the Lemma \ref{lem: l2_gain}, the next proposition formalizes an LMI-based condition guaranteeing the inequality \eqref{eq: hamilton-jacobi} to have the finite $\mathcal{L}_2$-gain of the mapping from $f^{(r)}$ to the fault estimation error.
	\begin{propos}[Finite $\mathcal{L}_2$-gain]
		Suppose there exist matrices $P\succ0, R,$ $Q$, $J$ and scalar $\rho \geq 0$ satisfying
		\begin{align}
			&\left[\begin{array}{ccc}
				L_{11} & (P + R C_a) D_a & X_{12} \\
				* & -{\rho}I  & \boldsymbol{0} \\
				* & * & -I
			\end{array}\right]\preceq0, \label{eq: l2_gain}\\
           &L_{11} := X+ \frac{1}{2} \bar{C}^{T} \bar{C},
			\label{eq: l11}
		\end{align}
and $X, X_{12}$ as defined in \eqref{eq: X_X12}, $\bar{C}$ in \eqref{eq: c_bar} and the remaining matrices in \eqref{eq: augmented_matrices}. Then, the $\mathcal{L}_2$-gain from $f^{(r)}$ to the fault estimation error $p = \bar{C}e $ in \eqref{eq: observer_dynamics_1} is upper bounded by $\sqrt{2 \rho}$.
		\label{propos: L_2_gain}
	\end{propos}
\emph{	\textbf{Proof}:}
	The proof is similar to the proof of Proposition \ref{propos: stability} and is given in Appendix \ref{ap: l_2_gain}.
	\hfill $\blacksquare$

Using Proposition \ref{propos: L_2_gain}, we next cast a semidefinite program where we seek to minimize the $\mathcal{L}_2$-gain from $f^{(r)}$ to $p=\bar{C}e$. We add the ISS LMI in \eqref{eq: stability_lmi} as a constraint to this program to enforce that the resulting filter also guarantees boundedness for bounded faults. The latter is important to avoid that the filter diverges (as the $\mathcal{L}_2$ criteria does not guarantee stability).
	
\begin{theorem}\textbf{\emph{(Optimal Fault Estimator)}}
Consider the augmented system dynamics \eqref{eq: augmented}, the filter \eqref{eq: nuio}, and the corresponding estimation error dynamics \eqref{eq: observer_dynamics_1}. To design the optimal fault estimator \eqref{eq: nuio}, solve the following convex program
		\begin{equation}
			\begin{array}{cl}
				\min \limits_{P, R, Q, J, \rho}  & \rho \\
				\text{s.t.} & 						\vspace{1 mm}
				\left[\begin{array}{cc}
					X+\epsilon I & X_{12} \\
					X_{12}^{T} & -I
				\end{array}\right] \preceq 0\\
				& \left[\begin{array}{ccc}
					L_{11} & (P + R C_a) D_a & X_{12} \\
					* & -{\rho}I  & \boldsymbol{0} \\
					* & * & -I
				\end{array}\right]\preceq 0 \\[5mm]

				& P \succ 0, \qquad \rho \geq 0,
			\end{array}
			\label{eq: mimization}
		\end{equation}
with given $\epsilon > 0$, $X, X_{12}$ as defined in \eqref{eq: X_X12}, $L_{11}$ in \eqref{eq: l11}, and the remaining matrices in \eqref{eq: augmented_matrices}. Let us denote the optimizers by $P^\star, R^\star, Q^\star, J^\star, \rho^\star,$ and based on \eqref{eq: variable_change}, define the matrices $E^\star = P^{\star^{-1}} R^\star , K^\star = P^{\star^{-1}} Q^\star$. Then, we have the following:
		\begin{enumerate}
			\item The ISS-gain from $f^{(r)}$ to $e$ is upper bounded by $ {2 \|P^\star (I+E^\star C_a) D_a\|}\epsilon^{-1}$.
			\item The $\mathcal{L}_2$-gain from $f^{(r)}$ to $\bar{C}e $ is upper bounded by $\sqrt{2 \rho^\star}$.
		\end{enumerate}
		\label{theorem: minimization}
	\end{theorem}
	\emph{\textbf{Proof}:}
	Theorem 1 follows from the above discussion, Proposition \ref{propos: stability}, and Proposition \ref{propos: L_2_gain}.
	\hfill $\blacksquare$\\
\begin{remark}[Perfect Estimation]
The developed methodology can guarantee zero estimation error for zero $f^{(r)}$, i.e., when the $r$-time derivative of the fault vector vanishes. This follows directly from the estimation error dynamics \eqref{eq: observer_dynamics_1} since if $f^{(r)} = \boldsymbol{0}$, the error dynamics becomes independent of external disturbances.
\end{remark}
	
	\begin{figure}[t!]
		\centering
		\smallskip
		\includegraphics[width=1\linewidth,keepaspectratio]{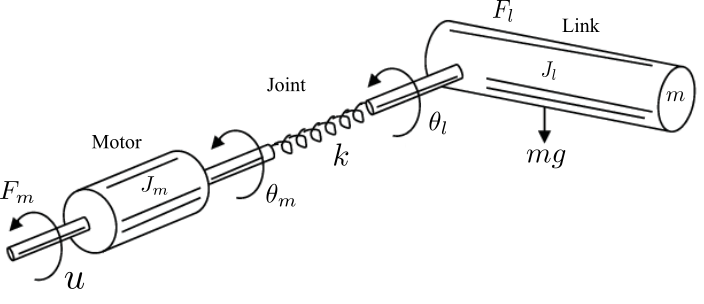}
		\caption{Benchmark System Schematic.}
		\label{fig: robot}
	\end{figure}
	\section{Simulation Results}  \label{sec: sim results}
	In this section, we evaluate the proposed method by a benchmark example for FDI \cite{Keliris2017, reppa2013adaptive,zhang2005sensor}. The dynamics of the system (a single-link robotic arm with a revolute elastic joint, see Fig. \ref{fig: robot} for a schematic) can be described as follows:
	\begin{equation*}
		\begin{aligned}
			& J_{l} \ddot{q}_{l}+F_{l} \dot{q}_{l}+k\left(q_{l}-q_{m}\right)+m g c \sin \left(q_{l}\right) =0, \\
			& J_{m} \ddot{q}_{m}+F_{m} \dot{q}_{m}-k\left(q_{l}-q_{m}\right) =k_{\tau} u,
		\end{aligned}
	\end{equation*}
	where ${q}_{l}$ and ${q}_{m}$ are the angular position of the link and the angular position of the motor, respectively. Constants $J_{l}$ and $J_{m}$ are the moments of inertia of the link and the motor and $F_{l}$ and $F_{m}$ are the viscous coefficients associated to friction acting at the link and the motor, respectively. The flexibility in the joint is modeled by a spring with a spring coefficient $k$, $m$ is the link mass, $g$ is the gravity constant, $c$ is the height of the link center of mass, $k_\tau$ is the amplifier gain, and $u$ is the torque input delivered by the motor. Units are in SI, and the parameters are: $ J_{l} = 4.5, J_{m} = 1, F_{l} = 0.5,
	F_{m} = 1, k = 2, m = 4, g = 9.8, c = 0.5,$ and $k_\tau = 1$. The torque input is $u = 2 \text{sin}(0.25t)$.
	
	By selecting $x_1 := \dot{q}_m, x_2 := q_m, x_3 := \dot{q}_l$, and $x_4 := q_l $, the system can be written in the form of \eqref{eq: sys_no_last_sensor}:

%\begin{figure}[b!] 	
%	\centering
%	\includegraphics[width=1\linewidth,keepaspectratio]{figs/estimated_fault_ob1_act_sin}
%	\caption{The sinusoidal actuator fault estimation error using the proposed method for $r=3$ and $r=5$.}
%	\label{fig: estimated_fault_ob1_act_sin}
%\end{figure}
	
	\begin{equation}
		\begin{aligned}
			\dot{x} &= A x+ B u+ S g\left(V x\right) + F_x f \\
			y &= C x + F_y f,
			\label{eq: benchmark_sys}
		\end{aligned}
	\end{equation}
	where $x := [x_1, x_2, x_3, x_4]^T$ is the state vector, and
	\begin{equation*}
		\begin{aligned}
			A &=\left[\begin{array}{cccc}
				-\frac{F_{m}}{J m} & -\frac{k}{J_{m}} & 0 & \frac{k}{J_{m}} \\
				1 & 0 & 0 & 0 \\
				0 & \frac{k}{J_{l}} & -\frac{F_{l}}{J_{l}} & -\frac{k}{J_{l}} \\
				0 & 0 & 1 & 0
			\end{array}\right], \\
				\end{aligned}
	\end{equation*}
	\begin{equation*}
	\begin{aligned}
			B&=\left[\begin{array}{cccc}
				\frac{k_{\tau}}{J_{m}} & 0 & 0 & 0 \\
			\end{array}\right]^T,
			&C&=\left[\begin{array}{cccc}
				0 & 1 & 0 & 0 \\
				0 & 0 & 0 & 1
			\end{array}\right], \\
			S &=\left[\begin{array}{cccc}
				0 & 0 & \frac{m g c}{J_{l}} & 0 \\
			\end{array}\right]^T,
			&V&=\left[\begin{array}{cccc}
				0 & 0 & 0 & 1 \\
			\end{array}\right]. \\
		\end{aligned}
	\end{equation*}

The nonlinearity is given by $g\left(Vx\right) =  - \sin \left(x_{4}\right)$, which is Lipschitz with constant $\alpha = 1$. We set initial conditions as $x(0) = [0.01,0.01,0.01,0.01]^T$. In addition, the first measurement is the angular position of the motor, and the second one is the angular position of the link. Then, we simulate for two different scenarios, one for the first sensor fault where we have
		\begin{equation*}
		\begin{aligned}
			F_x&=\left[\begin{array}{c}
				0 \\
				0  \\
				0  \\
				0  \\
			\end{array}\right],
			&F_y&=\left[\begin{array}{cc}
				 1  \\
				 0 \\
			\end{array}\right], \\
		\end{aligned}
	\end{equation*}
and the other one for the actuator fault with
		\begin{equation*}
	\begin{aligned}
		F_x&=\left[\begin{array}{c}
			1 \\
			0  \\
			0  \\
			0  \\
		\end{array}\right],
		&F_y&=\left[\begin{array}{cc}
			0  \\
			0 \\
		\end{array}\right]. \\
	\end{aligned}
\end{equation*}

	For both scenarios, we augment the benchmark system \eqref{eq: benchmark_sys} using \eqref{eq: augmented} with $r=1$. Next, we design an NUIO observer of the form \eqref{eq: nuio} by solving the minimization problem \eqref{eq: mimization} in Theorem 1, for the estimation of the actuator fault and the first sensor fault. To solve the convex program \eqref{eq: mimization}, we use the YALMIP toolbox in MATLAB.
	
		\begin{figure}[t!] 	
		\centering
		\smallskip
		\includegraphics[width=1\linewidth,keepaspectratio]{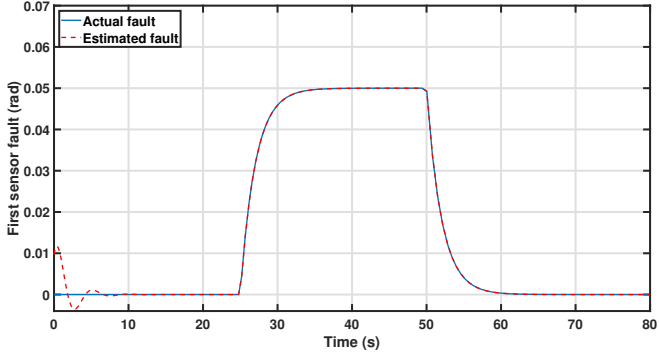}
		\caption{The actual sensor fault and its estimate using the proposed method.}
		\label{fig: estimated_fault_ob2_sen}
	\end{figure}

\begin{figure}[t!] 	
	\centering
	\smallskip
	\includegraphics[width=1\linewidth,keepaspectratio]{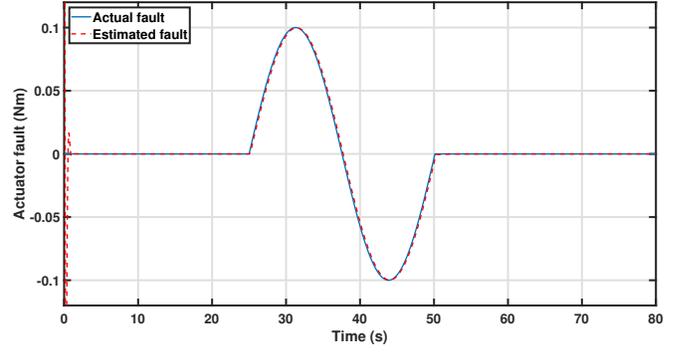}
	\caption{The actual actuator fault and its estimate using the proposed method.}
	\label{fig: estimated_fault_ob1_act}
\end{figure}

	Finally, we apply the designed NUIO to the system in the simulation. The initial condition of the observer in the simulation is taken as the zero vector. A gradually increasing (similar to an incipient fault \cite{Zhang}) sensor fault occurs at time $t = 25$  $(sec)$. Figure \ref{fig: estimated_fault_ob2_sen} depicts the estimated actuator fault and its actual value. For actuator fault, to indicate the capability of the developed method to estimate time-varying faults, a sinusoidal actuator fault with the same frequency of input is simulated (i.e., $f_x = 0.1 sin(0.25t)$). Figure \ref{fig: estimated_fault_ob1_act} shows the estimated actuator fault and its actual value. It can be seen that the estimated actuator and first sensor faults follow the actual fault with almost zero error in steady state.

	\section{Conclusion}\label{sec: conclusion}
	In this paper, a method for time-varying actuator and sensor faults estimation in nonlinear systems has been proposed. To this end, we have augmented the system dynamics with an approximated internal model for the fault vector (using ideas from ultra-local models). Then, a nonlinear unknown input observer is constructed for the augmented dynamics. The fault estimation problem has been re-formulated as a robust state estimation problem in the augmented state. We have provide sufficient conditions that ensure the stability of the observer error dynamics (guaranteeing asymptotic fault estimation) and robustness against model mismatch in the internal fault model (in terms of a finite $\mathcal{L}_2$-gain from fault model mismatch to fault estimation error). In Theorem \ref{theorem: minimization}, a convex minimization problem has been casted for destining optimal estimators. Finally, simulations for a benchmark system are given to illustrate the performance of the proposed methodology. The numerical simulations show that the proposed method properly estimates sensor and actuator faults for the benchmark system.

	\appendix
	\section{Appendices}
	\subsection{Lyapanov Function Proof} \label{ap: lyapanov}
	The inequality in \eqref{eq: lyapanov} can be derived as follows:
	\begin{equation}
		\begin{aligned} \dot{W} =& e^{T}\left(N^{T} P+P N\right) e+2 e^{T} P M S_a \delta g \\
			&- 2 e^{T} P M D_a f^{(r)}\\
			\leq & e^{T}\left(N^{T} P+P N\right) e+2\left\|e^{T} P M S_a \right\|\|\delta g\| \\
			&- 2 e^{T} P M D_a f^{(r)} \\
			\leq & e^{T}\left(N^{T} P+P N\right) e \\
			&+2\left\|e^{T} P M S_a\right\| \alpha\|(V_a  - J C_a) {e} \|- 2 e^{T} P M D_a f^{(r)} \\
			\leq & e^{T}\left(N^{T} P+P N\right) e +\alpha\big(\left\|e^{T} P M S_a\right\|^{2}+ \\
			&	\|(V_a - J C_a) e\|^{2} \big)- 2 e^{T} P M D_a f^{(r)}\\
			=& e^{T}\big(N^{T} P+P N+\alpha P M S_a S_a^{T} M^{T} P\\
			&+\alpha (V_a - J C_a)^{T} (V_a - J C_a)\big) e - 2 e^{T} P M D_a f^{(r)} \\
			:=& e^{T} \Delta e  - 2 e^{T} P M D_a f^{(r)}.
		\end{aligned}
		\label{eq: lyapanov_ap}
	\end{equation}
	
	\subsection{$\mathcal{L}_2$-Gain LMI} \label{ap: l_2_gain}
	If we define $W(e) :={e}^{T} P {e}$ with positive definite matrix $P$; then \eqref{eq: hamilton-jacobi} can be written as follows
	\begin{equation*}
		\begin{aligned}
			&e^{T}\left(N^{T} P+P N + \frac{1}{2} \bar{C}^{T} \bar{C}\right) e+2 e^{T} P M S_a \delta g \\
			&+ \frac{2}{\lambda^{2}} e^{T} P M D_a D_a^{T} M^{T} P e \leq 0. \\
		\end{aligned}
	\end{equation*}
	We can find an upper bound for the above inequality, similar to stability proof in Appendix \ref{ap: lyapanov}, as
	\begin{equation*}
		e^{T}(\Delta + \frac{1}{2} \bar{C}^{T} \bar{C} + \frac{2}{\lambda^{2}} P M D_a D_a^{T} M^{T} P) e  \leq 0,
	\end{equation*}
	where $\Delta$ is same as definition in \eqref{eq: lyapanov_ap}. Using Schur complement the above inequality is equivalent to
	\begin{equation*}
		\left[\begin{array}{cc}
			\Delta + \frac{1}{2} \bar{C}^{T} \bar{C}  & P M D_a \\
			* & -\frac{\lambda^2}{2}I
		\end{array}\right]\preceq 0.
	\end{equation*}
	If we follow the same procedure in proof of Proposition \ref{propos: stability}, the equivalent inequality can be found as
	\begin{equation*}
		\left[\begin{array}{ccc}
			X+ \frac{1}{2} \bar{C}^{T} \bar{C} &  (P + R C_a) D_a & X_{12} \\
			* & -\frac{\lambda^2}{2}I  & 0 \\
			* & * & -I
		\end{array}\right]\preceq0.
	\end{equation*}
	Finally, by defining a change of variable as $\rho := \frac{\lambda^2}{2}$, \eqref{eq: l2_gain} can be implied.
	
\section*{acknowledgment}
This publication is part of the project Digital Twin project
4.3 with project number P18-03 of the research programme
Perspectief which is (mainly) financed by the Dutch Research
Council (NWO).

	\bibliographystyle{IEEEtran}
	\bibliography{IEEEabrv,refs}

\begin{thebibliography}{10}
\providecommand{\url}[1]{#1}
\csname url@rmstyle\endcsname
\providecommand{\newblock}{\relax}
\providecommand{\bibinfo}[2]{#2}
\providecommand\BIBentrySTDinterwordspacing{\spaceskip=0pt\relax}
\providecommand\BIBentryALTinterwordstretchfactor{4}
\providecommand\BIBentryALTinterwordspacing{\spaceskip=\fontdimen2\font plus
\BIBentryALTinterwordstretchfactor\fontdimen3\font minus
  \fontdimen4\font\relax}
\providecommand\BIBforeignlanguage[2]{{%
\expandafter\ifx\csname l@#1\endcsname\relax
\typeout{** WARNING: IEEEtran.bst: No hyphenation pattern has been}%
\typeout{** loaded for the language `#1'. Using the pattern for}%
\typeout{** the default language instead.}%
\else
\language=\csname l@#1\endcsname
\fi
#2}}

\bibitem{ding2008model}
S.~X. Ding, \emph{Model-based fault diagnosis techniques: design schemes,
  algorithms, and tools}.\hskip 1em plus 0.5em minus 0.4em\relax Springer
  Science \& Business Media, 2008.

\bibitem{zhang2002robust}
X.~Zhang, M.~M. Polycarpou, and T.~Parisini, ``A robust detection and isolation
  scheme for abrupt and incipient faults in nonlinear systems,'' \emph{IEEE
  Transactions on Automatic Control}, vol.~47, no.~4, pp. 576--593, 2002.

\bibitem{hwang2009survey}
I.~Hwang, S.~Kim, Y.~Kim, and C.~E. Seah, ``A survey of fault detection,
  isolation, and reconfiguration methods,'' \emph{IEEE Transactions on Control
  Systems Technology}, vol.~18, no.~3, pp. 636--653, 2009.

\bibitem{Moradmand2020}
A.~Moradmand, B.~Shafai, and M.~Saif, ``A design procedure for robust actuator
  and sensor fault detection,'' in \emph{2020 7th International Conference on
  Control, Decision and Information Technologies (CoDIT)}, vol.~1.\hskip 1em
  plus 0.5em minus 0.4em\relax IEEE, 2020, pp. 709--714.

\bibitem{chen2012robust}
J.~Chen and R.~J. Patton, \emph{Robust model-based fault diagnosis for dynamic
  systems}.\hskip 1em plus 0.5em minus 0.4em\relax Springer Science \& Business
  Media, 2012, vol.~3.

\bibitem{narasimhan2008new}
S.~Narasimhan, P.~Vachhani, and R.~Rengaswamy, ``New nonlinear residual
  feedback observer for fault diagnosis in nonlinear systems,''
  \emph{Automatica}, vol.~44, no.~9, pp. 2222--2229, 2008.

\bibitem{esfahani2015tractable}
P.~Mohajerin~Esfahani and J.~Lygeros, ``A tractable fault detection and
  isolation approach for nonlinear systems with probabilistic performance,''
  \emph{IEEE Transactions on Automatic Control}, vol.~61, no.~3, pp. 633--647,
  2015.

\bibitem{de2001geometric}
C.~De~Persis and A.~Isidori, ``A geometric approach to nonlinear fault
  detection and isolation,'' \emph{IEEE Transactions on Automatic Control},
  vol.~46, no.~6, pp. 853--865, 2001.

\bibitem{liu2013sensor}
M.~Liu and P.~Shi, ``Sensor fault estimation and tolerant control for it{\^o}
  stochastic systems with a descriptor sliding mode approach,''
  \emph{Automatica}, vol.~49, no.~5, pp. 1242--1250, 2013.

\bibitem{liu2012fuzzy}
M.~Liu, X.~Cao, and P.~Shi, ``Fuzzy-model-based fault-tolerant design for
  nonlinear stochastic systems against simultaneous sensor and actuator
  faults,'' \emph{IEEE Transactions on Fuzzy Systems}, vol.~21, no.~5, pp.
  789--799, 2012.

\bibitem{Adaptive}
B.~Jiang, M.~Staroswiecki, and V.~Cocquempot, ``Fault accommodation for
  nonlinear dynamic systems,'' \emph{IEEE Transactions on Automatic Control},
  vol.~51, pp. 1578--1583, 2006.

\bibitem{Phan2021}
C.~P. Vo, H.~V. Dao, K.~K. Ahn, \emph{et~al.}, ``Robust fault-tolerant control
  of an electro-hydraulic actuator with a novel nonlinear unknown input
  observer,'' \emph{IEEE Access}, vol.~9, pp. 30\,750--30\,760, 2021.

\bibitem{zhu2015fault}
J.-W. Zhu, G.-H. Yang, H.~Wang, and F.~Wang, ``Fault estimation for a class of
  nonlinear systems based on intermediate estimator,'' \emph{IEEE Transactions
  on Automatic Control}, vol.~61, no.~9, pp. 2518--2524, 2015.

\bibitem{Patton_Input_Observability}
M.~HOU and R.~PATTON, ``Input observability and input reconstruction,''
  \emph{Automatica}, vol.~34, pp. 789--794, 1998.

\bibitem{van2020multiple}
C.~van~der Ploeg, M.~Alirezaei, N.~van~de Wouw, and P.~Mohajerin~Esfahani,
  ``Multiple faults estimation in dynamical systems: Tractable design and
  performance bounds,'' \emph{arXiv preprint arXiv:2011.13730}, 2020.

\bibitem{Flies_Ultra_Local}
M.~Fliess and C.~Join, ``Model-free control,'' \emph{International Journal of
  Control}, vol.~86, pp. 2228--2252, 2013.

\bibitem{sira2018active}
H.~Sira-Ram{\'\i}rez, A.~Luviano-Ju{\'a}rez, M.~Ram{\'\i}rez-Neria, and E.~W.
  Zurita-Bustamante, \emph{Active disturbance rejection control of dynamic
  systems: a flatness based approach}.\hskip 1em plus 0.5em minus 0.4em\relax
  Butterworth-Heinemann, 2018.

\bibitem{khalil2002nonlinear}
H.~K. Khalil, ``Nonlinear systems third edition,'' \emph{Patience Hall}, vol.
  115, 2002.

\bibitem{edwards2006comparison}
C.~Edwards and C.~P. Tan, ``A comparison of sliding mode and unknown input
  observers for fault reconstruction,'' \emph{European Journal of Control},
  vol.~12, no.~3, pp. 245--260, 2006.

\bibitem{edwards2000sliding}
C.~Edwards and S.~K. Spurgeon, ``A sliding mode observer based fdi scheme for
  the ship benchmark,'' \emph{European Journal of Control}, vol.~6, no.~4, pp.
  341--355, 2000.

\bibitem{frank1989robust}
P.~Frank and J.~Wunnenberg, ``Robust fault diagnosis using unknown input
  observer schemes,'' \emph{Fault Diagnosis in Dynamic Systems}, pp. 47--97,
  1989.

\bibitem{chen2006unknown}
W.~Chen and M.~Saif, ``Unknown input observer design for a class of nonlinear
  systems: an lmi approach,'' in \emph{American Control Conference}.\hskip 1em
  plus 0.5em minus 0.4em\relax IEEE, 2006, pp. 5--pp.

\bibitem{van19922}
A.~J. Van Der~Schaft, ``L 2-gain analysis of nonlinear systems and nonlinear
  state feedback $h_{\infty}$ control,'' \emph{IEEE Transactions on Automatic
  Control}, vol.~37, no.~6, pp. 770--784, 1992.

\bibitem{Keliris2017}
C.~Keliris, M.~M. Polycarpou, and T.~Parisini, ``An integrated learning and
  filtering approach for fault diagnosis of a class of nonlinear dynamical
  systems,'' \emph{IEEE Transactions on Neural Networks and Learning Systems},
  vol.~28, no.~4, pp. 988--1004, 2016.

\bibitem{reppa2013adaptive}
V.~Reppa, M.~M. Polycarpou, and C.~G. Panayiotou, ``Adaptive approximation for
  multiple sensor fault detection and isolation of nonlinear uncertain
  systems,'' \emph{IEEE Transactions on Neural Networks and Learning Systems},
  vol.~25, no.~1, pp. 137--153, 2013.

\bibitem{zhang2005sensor}
X.~Zhang, T.~Parisini, and M.~M. Polycarpou, ``Sensor bias fault isolation in a
  class of nonlinear systems,'' \emph{IEEE Transactions on Automatic Control},
  vol.~50, no.~3, pp. 370--376, 2005.

\bibitem{Zhang}
X.~Zhang, M.~M. Polycarpou, and T.~Parisini, ``A robust detection and isolation
  scheme for abrupt and incipient faults in nonlinear systems,'' \emph{IEEE
  Transactions on Automatic Control}, vol.~47, no.~4, pp. 576--593, 2002.

\end{thebibliography}

\end{document}